%% file: Marston_masterfile.tex
\documentclass[10pt,twoside,BCOR7mm,DIV15,headinclude,footexclude,
               cleardoubleempty,idxtotoc]
{scrartcl}

\usepackage{natbib}
\usepackage[font=small,labelfont=bf]{caption}
\usepackage[english]{babel}
\usepackage{graphicx}
\usepackage{hyperref}
\usepackage{scrpage2}
\usepackage{ifthen}
\usepackage{booktabs}
\usepackage{amsmath}
\usepackage{amssymb}
\usepackage{multicol}
\usepackage{float}
\usepackage{hyperref}

\hypersetup{breaklinks=true
,colorlinks=true,linkcolor=black,urlcolor=blue
,citecolor=black}

\addto\captionsenglish{%
}

\makeatletter
\renewcommand{\@biblabel}[1]{}
\renewcommand{\@cite}[2]{%
{#1\ifthenelse{\boolean{@tempswa}}{,#2}{}}}
\makeatother
\ofoot{\thepage}
\ifoot{}

\setcapindent{0em}
\setheadsepline{1pt}

\setkomafont{pagehead}{\normalfont\sffamily}
\setkomafont{pagenumber}{\normalfont\rmfamily}

\makeatletter
\newcommand{\listofcontributions}{\@starttoc{con}}

\newcommand{\l@contribution} {\@dottedtocline{1}{1.5em}{2.3em}}
\makeatother

\newenvironment{contribution}{
\setcounter{section}{0}
\setcounter{figure}{0}
\setcounter{table}{0}
}{
\newpage
\lehead{}
\rohead{}
}

\begin{document}

\setlength{\baselineskip}{2.5ex}

\begin{contribution}
\include{Marston_etal_Potsdam_WR_June2015}
\end{contribution}

%%-------------------------------------------------------

\end{document}

%% file: Marston_etal_Potsdam_WR_June2015.tex
% EXAMPLE AND TEMPLATE FILE FOR PROCEEDINGS OF THE WOLF-RAYET WORKSHOP.
% PLEASE REPLACE THE TEMPLATE TEXT BY YOUR OWN ARTICLE.
% NOTE THAT YOU MUST NOT PROCESS THIS FILE, BUT THE MASTER FILE:
% latex masterfile; dvips masterfile

% RUNNING AUTHOR: PUT AUTHOR NAMED HERE
\lehead{A.P.\ Marston, J.\ Mauerhan, P.\ Morris \& S.\ Van Dyk}

% RUNNING TITLE; SHORTEN THE TITLE IF NECESSARY
% IN CASE OF A ONE-PAGE CONTRIBUTION (POSTER),
% SQUEEZE AUTHORS AND TITLE IN THIS LINE (Author: Title ...)
\rohead{Marston et al: Finding WR Stars in The Milky Way}

\begin{center}
% FULL TITLE HEADING
{\LARGE \bf Finding Wolf-Rayet Stars in the Milky Way: Inputs to Star Formation and Stellar Evolution}\\
\medskip

% AUTHORS LIST
{\it\bf A.P.\ Marston$^1$, J.\ Mauerhan$^2$, P.\ Morris$^3$ \& S.\ Van Dyk$^3$}\\

% AFFILIATIONS
{\it $^1$European Space Astronomy Centre (ESAC), Spain}\\
{\it $^2$University of California Berkeley, California, USA}\\
{\it $^3$Infrared Processing and Analysis Center (IPAC), Caltech, California, USA}

% ABSTRACT
\begin{abstract}
The total population of Wolf-Rayet (WR) stars in the Galaxy is predicted by models to be as many as $\sim$6000 stars, and yet the number of catalogued WR stars as a result of optical surveys was far lower than this ($\sim$200) at the turn of this century. When beginning our WR searches using infrared techniques it was not clear whether WR number predictions were too optimistic or whether there was more hidden behind interstellar and circumstellar extinction. During the last decade we pioneered a technique of exploiting the near- and mid-infrared continuum colours for individual point sources provided by large-format surveys of the Galaxy, including 2MASS and Spitzer/GLIMPSE, to pierce through the dust and reveal newly discovered WR stars throughout the Galactic Plane. The key item to the colour discrimination is via the characteristic infrared spectral index produced by the strong winds of the WR stars, combined with dust extinction, which place WR stars in a relatively depopulated area of infrared colour-colour diagrams. The use of the Spitzer/GLIMPSE 8$\mu$m and, more recently, WISE 22$\mu$m fluxes together with cross-referencing with X-ray measurements in selected Galactic regions have enabled improved candidate lists that increased our confirmation success rate, achieved via follow-up infrared and optical spectroscopy. To date a total of 102 new WR stars have been found with many more candidates still available for follow-up. This constitutes an addition of $\sim$16\% to the current inventory of 642 Galactic WR stars. In this talk we review our methods and provide some new results and a preliminary review of their stellar and interstellar medium environments. We provide a roadmap for the future of this search, including statistical modeling, and what we can add to star formation and high mass star evolution studies.
\end{abstract}
\end{center}

% TEXT OF THE PAPER, TWO-COLUMN STYLE
\begin{multicols}{2}

\section{Motivation for finding Wolf-Rayet stars in the Milky Way}

At the start of our studies in 2003, we considered the distribution and numbers of Wolf-Rayet (WR) stars in the Milky Way. There were a number of motivating factors for finding finding more. 
A limited number of Wolf-Rayet stars were known, 227 in 2001, while around 20 times more than this were predicted to exist in the Galaxy (see \citet{marston:vdhucht01}. Sample groups of particular WR subtypes made evolutionary studies more difficult. WR to O star and WN versus WC sublass ratios are key for stellar evolution model tests. These values and their variation across the Galaxy were unclear. WR population estimates are major constraints for predicted evolutionary lifetimes (e.g., \citealt{marston:maeder14,marston:shenar14}). While studies of a larger sample of WR star ejecta nebulae could also provide information on evolutionary sequences and linkages to other high mass stars such as Luminous Blue Variables (LBVs).

Since 2001 more than 400 new WR stars have been discovered in the Galaxy. Predominantly due to new techniques in the infrared. A slew of estimates indicate a total number in the Galaxy as being between 1200 and 6000 stars \citep{marston:shara09, marston:rosslowe15}. More than 100 of these new discoveries come directly or indirectly from our work. We describe our methods and we indicate how our work is expected to continue in the future.

\section{Defining WR candidates}

\subsection{Spitzer/GLIMPSE survey candidates}
The best means of finding new stellar populations in the Galaxy is via infrared observations. The GLIMPSE legacy program on board the {\it Spitzer Space Observatory} allowed 4-colour broadband infrared imaging between 3.6 and 8.0$\mu$m covering the whole Galactic plane between galactic latitudes of -1 and +1 degrees. The extracted GLIMPSE point source catalogue and associated 2MASS sources provide the potential for identifying previously obscured and distant stellar populations \cite{marston:benjamin03}. It also provides an unbiased sample across the galactic plane, not just covering the dense massive star clusters where many WR stars have been found previously.

\begin{figure}[H]
\begin{center}
\includegraphics[width=\columnwidth]{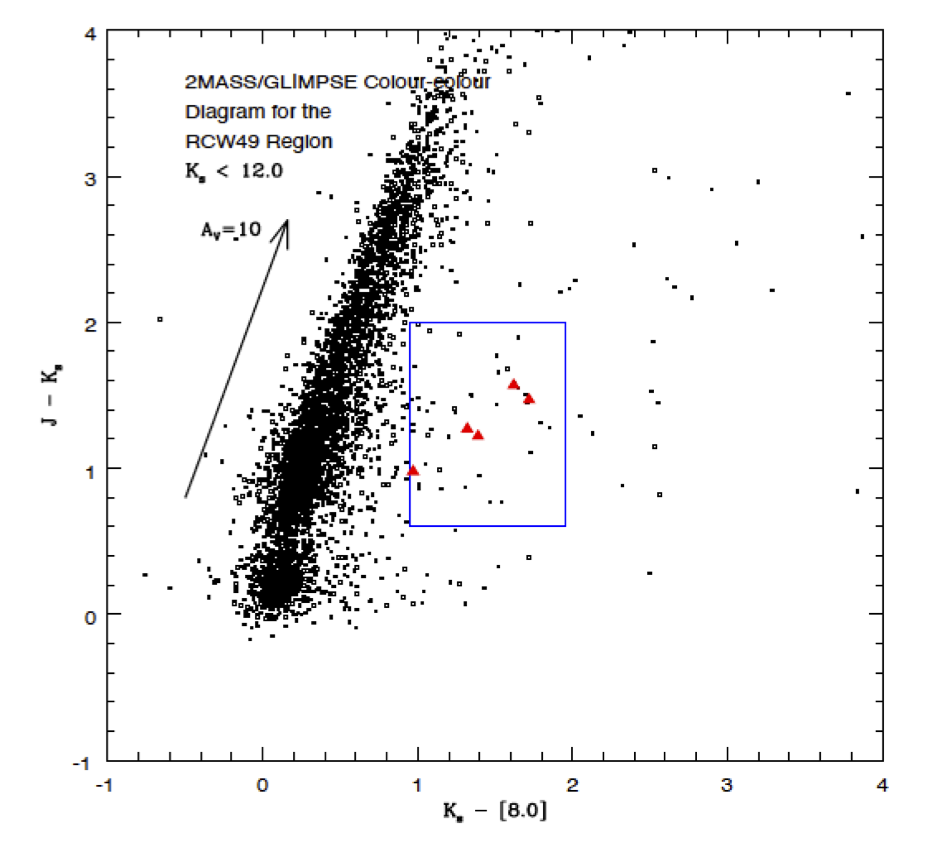}
\caption{Original colour-colour plot for objects in the RCW49 region for GLIMPSE identified objects with 2MASS K$_s$ band magnitudes $<$ 12.0, used in our NTT proposal. WR stars from the \citet{marston:vdhucht01} catalogue are shown as red triangles, while the blue box provides the boundary conditions used for identifying potential new WR candidates. A similar selection was done for 2x2 degree regions at galactic latitues of 312 and 320 degrees. The reddening vector is based on \citet{marston:indebetouw05}.
\label{marston:color-color plot}}
\end{center}
\end{figure}

Strong winds from WR stars are responsible for producing free-free emission which shows as an infrared excess in the spectral energy distribution (SED). \citet{marston:morris93} showed the observed wavelength spectral index of WR stars is $-2.95\pm0.25$, providing a shallower spectral index than that of a pure photosphere. With a limited range of spectral indices, there is a narrow range of possible infrared colours. Results from fields observed early in the GLIMPSE survey were used to distinguish candidate objects, including the field around Westerlund 2 and RCW49 (see Figure \ref{marston:color-color plot}). Candidate WR stars were defined by those objects in the colour-colour plot box shown in Figure \ref{marston:color-color plot} which contained all previously known WR stars. A similar candidate selection process was done for two less remarkable parts of the Galaxy at longitudes of 312 and 321 degrees.
%-----------One-column figure -----------------------------------
% Note that only the [H] option is allowed for placing 1-column figures!
Two main surprises of the work were, a. the fact that WR stars were found well away from stellar clusters; b. approximately 85\% of all candidates showed emission lines, most of which appear to be lower mass Be stars.

\subsection{Further WR discoveries}
\begin{figure}[H]
\begin{center}
\includegraphics[width=\columnwidth]{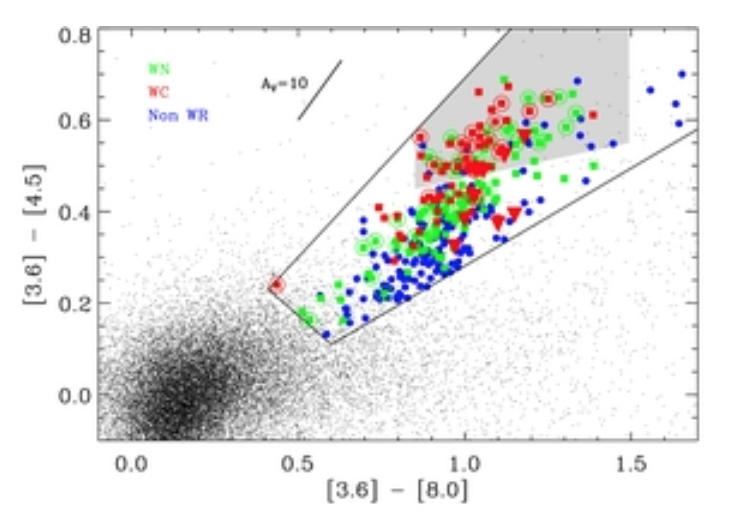}
\caption{WN and WC objects are seen in green and red in this colour-colour plot taken from \citet{marston:mauerhan11}. Candidate objects turning out not to be WR stars are shown blue. The gray shaded area indicates a 50\%+ rate of WR detection from the candidate list. Dusty WC stars are shown by inverted red triangles.
\label{marston:redeyes}}
\end{center}
\end{figure}

A larger followup of the galactic plane around l=312 degrees by \citet{marston:hadfield07} provided 15 more WR star discoveries, while \citet{marston:mauerhan11} provided a further 60 with somewhat improved candidate selection through GLIMPSE colour selections (see Figure \ref{marston:redeyes}). 
%-----------One-column figure -----------------------------------
% Note that only the [H] option is allowed for placing 1-column figures!
%-----------------------------------------------------------

%-----------One-column figure -----------------------------------
% Note that only the [H] option is allowed for placing 1-column figures!
\begin{figure}[H]
\begin{center}
\includegraphics[width=\columnwidth]{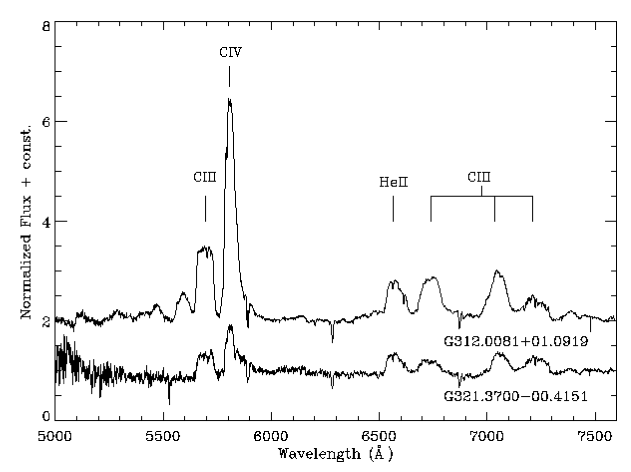}
\caption{CTIO/SOAR optical spectra taken by P.Morris and S.Van Dyk as confirmation to the SOFI near-infrared discoveries of WR60-5 and WR67-2.
\label{marston:WC_optical_spectra}}
\end{center}
\end{figure}
%-----------------------------------------------------------

\subsection{Optical and Near-Infrared Spectra of WR candidates}
Confirmation of WR candidates was achieved initially by near-infrared H and K band SOFI spectroscopy at the NTT on La Silla, Chile in 2004. Further optical spectroscopy of confirmed candidates was done in follow-up observations on the CTIO/SOAR telescope. Six WR stars have been discovered from this early work (see \citealt{marston:marston13}; \citealt{marston:roman11a}; \citealt{marston:roman11b}; \citealt{marston:roman12}). Optical spectra of two new WC stars found in this early work are shown in Figure \ref{marston:WC_optical_spectra}.

\subsection{Future candidate selection}
We are currently looking into improving candidate selection criteria based on near- and mid-infrared colours (including MIPSGAL/Spitzer and WISE fluxes). We can start to make knowledge-based estimates of the probability/confidence level of and object's classification based on its colours. In this way we can potentially determine a statistical value of WR numbers in the galaxy -- assuming a reasonable probability estimate is able to be done. To this end we have considered nearest-neighbour tests in infrared colour space. Potentially this could improve WR candidate identification considerably, but requires hypothesis testing via follow-up IR spectroscopic observations that we expect to make in the summer of 2015.

\section{Followup of new WR stars - origins of isolated WR stars}

Of the more than 100 WR stars discovered in follow-up spectroscopy of infrared colour-selected candidates, a large fraction of them are found well away from stellar clusters. In general, we have a bias against objects being found in clusters due to Spitzer confusion, our work therefore complements surveys such as those done in Westerlund 1 \citep{marston:crowther06}. A number of new WR stars have been found towards the edges of stellar clusters (e.g. see \citealt{marston:roman11a}; and the Danks clusters objects in \citealt{marston:mauerhan11}). This hints at significant numbers of runaway stars from central dense stellar clusters.

Further out in the galaxy, a number of apparently isolated WR stars may well be scattered out from stellar clusters at earlier times \citep[e.g., see] []{marston:mackey13} and indeed 25\% of local O stars may be runaways \citep{marston:blaauw93}. GAIA results will help with determining proper motions of WR stars in the future.

But some objects appear too far from star forming areas to have been scattered from high density star forming sites within an average WR lifetime and suggest the possibility of {\it in situ} star formation, possibly associated with looser stellar associations. Although the lack of stellar clusters may simply be due to elusive lower mass clusters lost in infrared  galactic confusion -- ``false negative clusters'' \cite{marston:hanson10}.

\section{Conclusions}

Broad-band infrared colours have been successful in finding obscured and distant WR stars in our Galaxy. We plan to extend the work to better provide numbers and distributions of WR stars and their subclasses. Such information helps to constrain stellar evolutionary models of high mass stars and the current assumptions they make.

An alternative narrow-band infrared imaging technique, centred on the wavelengths of key spectral line features in WR spectra, has been performed in the Galactic plane and new WR discoveries have been reported in several papers including \citet{marston:shara09} and \citet{marston:faherty14}.

Our studies show that WR stars are not all grouped in dense stellar clusters but many may be runaways or possibly created in relatively isolated regions or loose associations. Many of the objects also have nebulae (including a number of apparent ejecta nebulae) associated with them which provide a better statistical basis for the study of WR subclasses and ejecta plus timing of ejecta events.

%----------- Double-column figure -----------------------------------
%\begin{figure*}[!t]
%\begin{center}
%\includegraphics
%  [width=\textwidth]{wr2015-paper-twocol}
%\caption{A figure that stretches over both 
%columns. This figure environment
%is a so-called ``starred'' environment (\texttt{figure*}).
%\label{example:bigfig}}
%\end{center}
%\end{figure*}
%--------------------------------------------------------------------

\bibliographystyle{aa} % style aa.bst
\bibliography{Marston_etal_Potsdam_WR_June2015}

\end{multicols}